\documentclass[12pt]{article}
\pdfoutput=1

\usepackage{draft} 
\usepackage{hyperref}
\usepackage{graphicx,color,subfig}
\usepackage{cite}
\usepackage{mciteplus}
\usepackage{skak}
\DeclareFontFamily{OT1}{pzc}{}
\DeclareFontShape{OT1}{pzc}{m}{it}{<-> s * [1.10] pzcmi7t}{}
\DeclareMathAlphabet{\mathpzc}{OT1}{pzc}{m}{it}
\usepackage{mathtools}

\numberwithin{equation}{section}

\def\0{{(0)}}
\def\1{{(1)}}
\def\2{{(2)}}

\def\<{\langle }
\def\>{\rangle }

\def\eads${H$_3$}

\def\({\left (}
\def\){\right )}
\def\[{\left [}
\def\]{\right ]}
\def\d{\mathrm{d}}
\def\<{\langle }
\def\>{\rangle }

\def\beq{\begin{equation}\begin{aligned}}
\def\eeq{\end{aligned}\end{equation}}

\def\be#1\ee{\begin{align}#1\end{align}}

\begin{document}

\unitlength = .8mm

\begin{titlepage}

\begin{center}

\hfill \\
\hfill \\
\vskip 1cm

\title{Conformal Basis, Optical Theorem,\\
and the Bulk Point Singularity}

\author{Ho Tat Lam$^1$ and Shu-Heng Shao$^2$}

\address{
$^1$Department of Physics, Princeton University,\\
Princeton, NJ 08540, USA\\
$^2$School of Natural Sciences, Institute for Advanced Study, \\Princeton, NJ 08540, USA
}

\end{center}

\vspace{2.0cm}

\begin{abstract}

We study general properties of the conformal basis, the space of  wavefunctions in  $(d+2)$-dimensional Minkowski space that are primaries of the Lorentz group $SO(1,d+1)$.  Scattering amplitudes written in this basis have the same symmetry as $d$-dimensional conformal correlators.  
We translate  the optical theorem, which is a direct consequence of unitarity, into the conformal basis.  In the particular case of a tree-level exchange diagram, the optical theorem  takes the form of a conformal block decomposition on the principal continuous series, with OPE coefficients being the three-point coupling written in the same basis.  
We further discuss the relation between the massless conformal basis and the bulk point singularity in AdS/CFT.  
Some three- and four-point amplitudes in (2+1) dimensions are explicitly computed in this basis to demonstrate these results.

\end{abstract}

\vfill

\end{titlepage}

\eject

\tableofcontents

\section{Introduction}

The Lorentz group of  $(d+2)$-dimensional Minkowski space is the same as the Euclidean conformal group in $d$ dimensions.  
This makes it possible to interpret a $(d+2)$-dimensional scattering amplitude as a conformal correlator in $d$ dimensions. 
Recently,  building on the earlier work of \cite{deBoer:2003vf},  a basis of flat space wavefunctions has been constructed in \cite{Cheung:2016iub,Pasterski:2016qvg,Pasterski:2017kqt}, where  scattering amplitudes in $\mathbb{R}^{1,d+1}$ take the form of $d$-dimensional conformal correlators.  This basis, called the conformal primary basis, or simply the \textit{conformal basis}, serves as a natural basis for the  study of two-dimensional conformal symmetries in four-dimensional flat space scattering amplitudes \cite{Banks:2003vp,Barnich:2009se,Barnich:2010eb,Barnich:2011ct,Strominger:2013lka,Kapec:2014opa,He:2015zea,Lipstein:2015rxa,Cardona:2015woa,Cheung:2016iub,Kapec:2016jld,He:2017fsb,Strominger:2017zoo,Nande:2017dba} (see \cite{Kapec:2017gsg} for discussions in general dimensions).

More explicitly,  we consider scalar wavefunctions in $\mathbb{R}^{1,d+1}$ that transform as $d$-dimensional conformal primaries under $SO(1,d+1)$ constructed in \cite{deBoer:2003vf,Cheung:2016iub,Pasterski:2016qvg,Pasterski:2017kqt}.  These wavefunctions, called the \textit{conformal primary wavefunctions}, are labeled by a conformal dimension $\Delta$ and a point $\vec x\in\mathbb{R}^d$, rather than an on-shell momentum in $\mathbb{R}^{1,d+1}$.  Consequently, scattering amplitudes of these wavefunctions are functions of $\Delta_i,\vec x_i$ and transform covariantly as $d$ dimensional conformal correlators under $SO(1,d+1)$. 
 Through the study of their inner products, it was further shown in \cite{Pasterski:2017kqt} that the continuum of conformal primary wavefunctions with $\Delta\in \frac d2 + i \mathbb{R}$ forms a basis of normalizable solutions to the wave equation.  This range of the conformal dimension is known as the principal continuous series of unitary irreducible representations of $SO(1,d+1)$, which plays an important role in the study of conformal field theory (CFT) (see, for example, \cite{Costa:2012cb,Gadde:2017sjg,Hogervorst:2017sfd,Caron-Huot:2017vep,Simmons-Duffin:2017nub}).

 In this paper we further explore  general properties of  scattering amplitudes  in the conformal basis.  
 One interesting question is the implication of unitarity of the $\mathcal{S}$-matrix in this basis.  
We approach this question by translating  the optical theorem, which is a direct consequence of unitarity, into the conformal basis.  
 In the case of a tree-level massive scalar exchange diagram, the optical theorem in the conformal basis takes the form of a conformal block decomposition on the principal continuous series:\footnote{In taking the imaginary part of the four-point function $f(z,\bar z)$, we have assumed all the conformal dimensions are analytically continued to be real.}
\begin{align}
 \text{Im}\, f(z,\bar z) = \pi m^d\, \int_{-\infty}^\infty d\nu \mu(\nu)\, C\left(\Delta_1,\Delta_2; \frac d2+i\nu\right) C\left( \Delta_3,\Delta_4 ; \frac d2 -i \nu\right)
\Psi_{\frac d2+i\nu}(z,\bar z)\,,
\end{align}
where $f(z,\bar z )$ is the four-point amplitude in the conformal basis and  $z,\bar z$ are the cross ratios.  $m$ is the mass of the intermediate particle.   $C(\Delta_1,\Delta_2;\Delta)$ is the  coefficient of the three-point amplitude written in the conformal basis.  $\mu(\nu)$ is a measure factor given in \eqref{measure}.  Finally, $\Psi_\Delta(z,\bar z )$ is the shadow-symmetric conformal partial wave \cite{Ferrara:1972uq,Dolan:2000ut,Dolan:2011dv,SimmonsDuffin:2012uy}.   The derivation of this conformal block decomposition follows from the completeness relation of the conformal primary wavefunctions on the principal continuous series $\Delta\in \frac d2+i \mathbb{R}$.  The final expression is very reminiscent of the split representation for  Witten diagrams in AdS \cite{Penedones:2010ue,Costa:2014kfa}.  

To verify the above optical theorem in concrete examples, we  consider scalar scattering amplitudes in (2+1) spacetime dimensions with a cubic coupling. The corresponding conformal correlators are one-dimensional with $SL(2,\mathbb{R})$ covariance. 
The three-point function takes the form of a standard CFT three-point function with coefficient $C(\Delta_1,\Delta_2;\Delta_3)$ given in terms of the gamma functions:
\begin{align}
g\frac{m^{\Delta_1+\Delta_2-3}}{2^{\Delta_1+\Delta_2}}
\frac{\Gamma\(\frac{\Delta_1+\Delta_3-\Delta_2}{2}\)\Gamma\(\frac{\Delta_2+\Delta_3-\Delta_1}{2}\)}{\Gamma(\Delta_3)}
\frac{1}{|x_{12}|^{\Delta_1+\Delta_2-\Delta_3}|x_{13}|^{\Delta_1+\Delta_3-\Delta_2}|x_{23}|^{\Delta_2+\Delta_3-\Delta_1}} \,.
\end{align}
  The four-point function with identical external dimension $\Delta_\phi$ also takes a particularly simple form 
\begin{align}
 f(z) = 
\mathcal{N}_{\Delta_\phi}\,
{z\over \sqrt{z-1}}
\left[ \,
e^{i \pi (2\Delta_\phi-\frac 32)}
+z^{2\Delta_\phi-\frac 32}
+\left( {z\over z-1}\right)^{2\Delta_\phi-\frac 32}
\, \right]\,,
\end{align}
where $z>1$ is the real cross ratio\footnote{Recall that in one dimension there is only one independent cross ratio of a four-point function.  Here the four-point function $f(z)$ (what we call $f_{12\leftrightarrow 34}(z)$ in the main text) is computed in the crossing channel where particle 1 and 2 are incoming while 3 and 4 are outgoing.  The other ranges of the cross ratio on the real line are realized by the other two crossing channels.} parametrizing the scattering angle and $\Delta_\phi$ is the conformal dimension we assign to the four external particles.  $\mathcal{N}_{\Delta_\phi}$ is a normalization constant given in the main text.  We show that the imaginary part of this four-point function can indeed be expanded on the conformal partial waves with coefficients being  $C(\Delta_\phi,\Delta_\phi;d/2+i\nu)C(\Delta_\phi,\Delta_\phi; d/2-i\nu)$.  We further discuss the implication of crossing symmetry of the two-to-two scattering amplitudes in the conformal basis.

Various properties of the conformal basis have been explored recently.  In \cite{Cheung:2016iub} the soft photon and graviton theorems are studied in the conformal basis in $(3+1)$ spacetime dimensions.  The massive scalar three-point amplitude is shown to be equal to the standard scalar CFT three-point function in the special mass limit in \cite{Pasterski:2016qvg}.  The tree-level gluon low-point amplitudes in the conformal basis have been computed in \cite{Pasterski:2017ylz}.  
The BCFW relation \cite{Britto:2004ap,Britto:2005fq} in this basis and its potential interpretation as the conformal block decomposition were explored in \cite{Pasterski:2017ylz,block}.  The factorization singularity has also been investigated in \cite{Cardona:2017keg,NVWZ}.

We then  turn to the relation between the \textit{massless} conformal basis in $\mathbb{R}^{1,d+1}$ and the bulk point singularity in $AdS_{d+2}/CFT_{d+1}$ \cite{Gary:2009ae,Heemskerk:2009pn,Penedones:2010ue,Okuda:2010ym,Maldacena:2015iua}.   The bulk point singularity is a singularity of perturbative holographic correlators in AdS/CFT that arises from Landau diagrams in the bulk.  It has been used to probe the flat space limit of AdS/CFT \cite{Susskind:1998vk,Polchinski:1999ry,Giddings:1999jq,Fitzpatrick:2011ia,Fitzpatrick:2011jn,Fitzpatrick:2011hu,Fitzpatrick:2011dm} and diagnose   bulk locality.  
We discuss how   the bulk point singularity of a Witten diagram in $AdS_{d+2}$, under certain assumptions, is computed by the same amplitude in the massless conformal basis in $\mathbb{R}^{1,d+1}$.  

In the example of scalar four-point amplitudes in (2+1) dimensions, the relation to the bulk point singularity in $AdS_3/CFT_2$ suggests that the one-dimensional correlators in the conformal basis should be interpreted as two-dimensional Lorentzian correlators, restricted to  the configuration with real cross ratio.  In Appendix \ref{app:2d} we present such a candidate  $2d$ Euclidean four-point function whose Lorentzian versions, when restricted to the bulk point singularity configuration, reproduce these one-dimensional correlators from different crossing channels.  We further show that this $2d$ correlator satisfies the crossing equation and  has a positive $SL(2,\mathbb{C})$ block decomposition with a simple spectrum of single-trace and double-trace intermediate operators.   The physical origin of this $2d$ extension remains to be understood.

This paper is organized as follows.  In Section \ref{sec:basis} we review both the massive and massless conformal bases.  In Section \ref{sec:1dcorrelator} we present explicit results for the three- and four-point correlators in a simple scalar (2+1)-dimensional model. In Section \ref{sec:optical} we translate the optical theorem into the conformal basis in general spacetime dimensions, and verify it explicitly for the scalar model in (2+1) dimensions. In Section \ref{sec:bulkpoint} we discuss the relation between the massless conformal basis and the bulk point singularity in AdS/CFT. 
 In Appendix \ref{app:2d}, we consider a $2d$ extension of the $1d$ correlators for the scalar model considered above.

\section{Conformal Primary Bases}\label{sec:basis}

In this section we review scalar conformal primary wavefunctions introduced in \cite{deBoer:2003vf,Cheung:2016iub,Pasterski:2016qvg,Pasterski:2017kqt}.  The construction of these wavefunctions in flat space proceeds naturally through the embedding space formalism in CFT \cite{Dirac:1936fq,Mack:1969rr,Cornalba:2009ax,Weinberg:2010fx,Costa:2011mg,Costa:2011dw,Costa:2014kfa}.  

The flat space coordinates of $\mathbb{R}^{1,d+1}$ will be denotes by $X^\mu$ with $\mu=0,1,\cdots, d+1$.  Our convention on the spacetime signature is $(-+\cdots +)$. We will parametrize an outgoing/incoming null momentum $k^\mu$ in $\mathbb{R}^{1,d+1}$ as
\begin{align}\label{qmap}
k^\mu  = \pm \omega \,q^\mu (\vec x) \equiv \pm\omega\, ( 1+|\vec x|^2 \,,\, 2\vec x \,,\, 1-|\vec x|^2)\,,
\end{align}
where $\vec x\in\mathbb{R}^d$ labels the direction of the null momentum and $\omega>0$ is a scale.  On the other hand, an outgoing/incoming  timelike momentum will be parametrized in terms of $y>0$ and $\vec z\in\mathbb{R}^d$ as
\begin{align}\label{pmap}
p^\mu = \pm m \hat p(y,\vec z) \equiv  \pm  m\left({1+y^2+|\vec z|^2\over 2y} \,,\, {\vec z\over y}\,,\,{1-y^2-|\vec z|^2\over 2y}\right)\,.
\end{align}
Note that $\hat p^2 =-1$.

Scattering amplitudes are usually written in the basis of plane waves $e^{\pm i k^\mu  X_\mu}$ which are eigenfunctions of translations.    In this paper we consider an alternative basis of wavefunctions $\varphi^\pm_\Delta(X^\mu;\vec x)$ that are labeled by a ``conformal dimension" $\Delta $ and a point $\vec x\in\mathbb{R}^d$, instead of an on-shell momentum in $\mathbb{R}^{1,d+1}$.  The $\pm$ superscript distinguishes an outgoing ($+$) wavefunction from an incoming ($-$) one. 
Conformal primary wavefunctions are defined such that, under a  Lorentz group $SO(1,d+1)$ transformation, the wavefunction $\varphi^\pm_\Delta(X^\mu;\vec x)$  transforms covariantly  as a scalar conformal primary operator in $d$ spacetime dimension:
\begin{align}\label{covariance}
\varphi_\Delta \left( \Lambda^\mu_{~\nu} X^\nu ;\vec x\,'(\vec x) \right)
= \left|  {\partial \vec x\,' \over \partial \vec x}\right|^{-\Delta/d} \,
\varphi_\Delta( X^\mu ;  \vec x)\,,
\end{align}
where  $\vec x\,'(\vec x)$ is a non-linear $SO(1,d+1)$ transformation  on $\vec x\in\mathbb{R}^d$ and $\Lambda^\mu_{~\nu}$ is the associated group element in the $(d+2)$-dimensional representation.

In the massless case, the conformal primary wavefunction $\varphi^\pm_\Delta(X^\mu;\vec x)$ can be easily written down \cite{deBoer:2003vf,Cheung:2016iub,Pasterski:2016qvg,Pasterski:2017kqt,Campiglia:2015lxa,Campiglia:2017dpg}:
\begin{align}\label{masslessCPW}
\varphi^\pm_\Delta(X^\mu;\vec x) =\mathcal{N} \, {1\over (-q(\vec x) \cdot X \mp i\epsilon)^\Delta}\,,
\end{align}
where we have introduced an $i\epsilon $ prescription to circumvent the singularity on the lightsheet $q\cdot X=0$. 
Here $\mathcal{N}={(\mp i)^\Delta \Gamma(\Delta)}$ is a normalization constant we choose for later convenience.  
The massless conformal primary wavefunction can be expanded on the plane waves via a Mellin transform of the scale $\omega$ in \eqref{qmap}:
\begin{align}
\varphi_\Delta^\pm(X^\mu;\vec x) = \int_0^\infty d\omega \,\omega^{\Delta-1}\, e^{\pm i\omega q\cdot X -\epsilon \omega} \,.
\end{align}
In \cite{Pasterski:2017kqt} it was shown that the continuum of  conformal primary wavefunctions on the $\Delta\in \frac d2 +i\mathbb{R}$ spans a complete set of delta-function-normalizable solutions (with respect to the Klein-Gordon inner product) to the massless Klein-Gordon equation.\footnote{There is another basis of massless conformal primary wavefunctions that is the shadow of \eqref{masslessCPW}.  We will not discuss this shadow basis in this paper.}
  This range of $\Delta$ is known as the \textit{principal continuous series} of $SO(1,d+1)$.

Let us now proceed to the massive case.  Similar to the massless case, we define a massive scalar conformal primary wavefunction $\phi_\Delta^\pm(X^\mu;\vec x)$ as a solution to the massive Klein-Gordon equation of mass $m$ in $\mathbb{R}^{1,d+1}$ that transforms covariantly as \eqref{covariance} under the Lorentz group $SO(1,d+1)$.  
  We can always expand an outgoing/incoming solution $\phi_\Delta^\pm(X^\mu;\vec x)$ to the massive Klein-Gordon equation on the plane waves as \cite{Pasterski:2016qvg}:
\begin{align}\label{massiveCPW}
\phi_\Delta^\pm(X^\mu;\vec x) = \int[d\hat p]\, G_\Delta(\hat p;\vec x)\,e^{\pm i m \hat p \cdot X}\,,
\end{align}
with some Fourier coefficient $G_\Delta(\hat p;\vec x)$.  Here $\int [ d\hat p]$ is  a Lorentz invariant integral over all the outgoing unit timelike vectors, which form a copy of two-dimensional hyperbolic space $H_{d+1}$:
\begin{align}
H_{d+1}:~ - (\hat p^0)^2  +\sum_{i=1}^{d+1} (\hat p^i)^2= -1\,,~~\hat p^0>0\,.
\end{align}
  We can write this measure $[d\hat p]$ more explicitly in terms of the hyperbolic coordinates $(y,\vec x)$ in \eqref{pmap} as
\begin{align}
\int [d\hat p] = \int_0^\infty {dy\over y^{d+1}} \int_{\mathbb{R}^d}d^d\vec z
= \int {d^{d+1}\hat p\over \hat p^0}
\,.
\end{align}
It now remains to determine the Fourier coefficient $G_\Delta(\hat p;\vec x)$.  Requiring the conformal covariance \eqref{covariance} of $\phi^\pm_\Delta(X^\mu;\vec x)$, the Fourier coefficient is determined to be the scalar bulk-to-boundary propagator  in the $(d+1)$-dimensional hyperbolic space $H_{d+1}$ \cite{Witten:1998qj}:
\begin{align}
G_\Delta(y,\vec z;\vec x) = \left( {y\over y^2+|\vec z-\vec x|^2}\right)^\Delta\,.
\end{align}
Similar to the massless case,  it was shown in \cite{Pasterski:2017kqt}  that the continuum of massive conformal primary wavefunctions on $\Delta\in \frac d2 +i \mathbb{R}_{\ge0}$ spans a complete set of normalizable solutions to the massive Klein-Gordon equation.



So far we have been talking about the wavefunction, but the above discussion can be immediately carried over to arbitrary scattering amplitudes in $\mathbb{R}^{1,d+1}$.  Consider an $n$-point scattering amplitude\footnote{The amplitude $\mathcal{T}$ is related to the connected part of the $\mathcal{S}$-matrix as $S_{conn} = i(2\pi)^{d+2} \mathcal{T}(k_i) \delta^{(d+2)}(\sum_i k_i)$ in $\mathbb{R}^{1,d+1}$.}
 $\mathcal{T}(k_\ell,p_j)\delta^{(d+2)} (\sum_\ell k_\ell +\sum_j p_j)$ of  scalars in momentum space  where $k_\ell$ and $p_j$ are the null and timelike momenta, respectively, for the external particles.  This amplitude can be transformed into the conformal primary basis via a Mellin transform for each massless external null momentum and an integral over $H_{d+1}$ \eqref{massiveCPW} for each massive external momentum:
\begin{align}\label{changebasis}
&\mathcal{\widetilde A}(\Delta_i ,\vec x_i ) \equiv\\
&\left(  \prod_{\ell:\,\text{massless}} \int_0^\infty d\omega_\ell \, \omega_\ell^{\Delta_\ell-1}\,\right)
\left(
\prod_{j:\,\text{massive}}
\int [d\hat p_j] G_{\Delta_j }(\hat p_j ;\vec x_j)\,
\right)
\mathcal{T}(k_\ell,p_j)\delta^{(d+2)} (\sum_\ell k_\ell +\sum_j p_j)\,,\notag
\end{align}
where $k^\mu  = \pm \omega q^\mu(\vec x)$ and $p^\mu = \pm m \hat p^\mu$.    Due to the conformal covariance of the conformal primary wavefunctions \eqref{covariance}, the amplitude $\mathcal{\widetilde A}(\Delta_i ,\vec  x_i)$ in the conformal basis is guaranteed to transform like a $d$-dimensional conformal correlator of scalar primaries with conformal dimensions $\Delta_i$  under $SO(1,d+1)$:
\begin{align}
\mathcal{\widetilde A}\left(\Delta_i ,\vec x\,'_i (\vec x_i)\right)
= \left( \prod_{k=1}^n\left| {\partial \vec x\,'_k \over \partial \vec x_k}\right|^{-\Delta_k/d} \right)\, \mathcal{\widetilde A}(\Delta_i , \vec x_i)\,.
\end{align}

\section{One-Dimensional Conformal Correlators}\label{sec:1dcorrelator}

In this section we consider conformal bases in (2+1) spacetime dimensions.  The amplitudes in the conformal basis take the form of  one-dimensional conformal correlators with $SL(2,\mathbb{R})$ symmetry.  This is the simplest nontrivial spacetime dimension where the resulting correlators are simple to analyze.  

\subsection{Three-Point Function}\label{sec:3pt}


Consider a perturbative theory in (2+1) dimensions consisting of one real massless scalar field $\Phi$ and one real massive scalar $\Phi_m$ of mass $m$, interacting through a cubic vertex $g\Phi^2 \Phi_m$.\footnote{Computationally, the change of basis integral is usually easier for the massless conformal basis than the massive one. However,  the three-point amplitude with all massless particles suffers from either an UV or IR divergence (depending on the conformal dimensions) in the change of basis integral.  We will hence consider the next simplest case where there is one massive particle, whose mass regulates the divergence, and two massless particles in (2+1) dimensions.} 
In momentum space, the tree-level three-point  amplitude of a massive scalar  with momentum $p^\mu = -m\hat p^\mu$ decaying into a pair of massless scalar  with momenta $k_\ell^\mu= \omega_\ell q^\mu(x_\ell)$ ($\ell=1,2$) is
\begin{align}
 i \mathcal{T}_3  =i g\,,
\end{align}
where $g$ is the three-point coupling. 
Using \eqref{changebasis}, the three-point amplitude written in the conformal basis is
\beq
\mathcal{\widetilde A}_3(\Delta_i,x_i)&=g\prod_{\ell=1}^{2}\int_0^{\infty}\omega_\ell^{\Delta_\ell-1}d\omega_\ell\int_0^\infty \frac{d y}{y^2}\int_{-\infty}^\infty dz\ \(\frac{y}{y^2+|z-x_3|^2}\)^{\Delta_3} \notag\\
&\times \,\delta^{(3)}(\omega_1 q^\mu(x_1)+\omega_2 q^\mu(x_2)- m \hat p^\mu )\,.
\eeq
The $\delta$-function can be used to localized the integrals in $y$, $z$ and $\omega_2$:
\beq
\delta^{(3)}(\omega_1 q^\mu(x_1)+\omega_2 q^\mu(x_2)- m \hat p^\mu )=\frac{y^{3}(\omega_1+\omega_2)}{m^{2}\omega_1|x_1-x_2|^2}\delta\(z-z^*\)\delta\(y-y^*\)\delta\(\omega_2-\omega_2^*\)\,,
\eeq
where
\beq
z^*=\frac{\omega_1x_1+\omega_2x_2}{\omega_1+\omega_2},\quad y^*=\frac{m}{2(\omega_1+\omega_2)},\quad \omega_2^*=\frac{m^2}{4\omega_1|x_1-x_2|^2}\,.
\eeq
The remaining integration in $\omega_1$ is
\beq
\widetilde{\mathcal{A}}_3(\Delta_i,x_i)=\frac{g m^{2\Delta_2+\Delta_3-3}}{2^{2\Delta_2-\Delta_3-1}|x_1-x_2|^{2\Delta_2-2\Delta_3}}\int_0^\infty\d\omega_1\
\frac{\omega_1^{\Delta_1-\Delta_2+\Delta_3-1}}{\(m^2|x_2-x_3|^2+4|x_1-x_2|^2|x_1-x_3|^2\omega_1^2\)^{\Delta_3}} \,.
\eeq
The integration converges if Re$\(\Delta_1-\Delta_2-\Delta_3\right)<0$ and Re$\(\Delta_1-\Delta_2+\Delta_3\right)>0$. The final three-point function takes the form of a standard three-point function in an one-dimensional conformal theory
\beq
\widetilde{\mathcal{A}}_3(\Delta_i,x_i)=\frac{C(\Delta_1,\Delta_2;\Delta_3)}{|x_1-x_2|^{\Delta_1+\Delta_2-\Delta_3}|x_1-x_3|^{\Delta_1+\Delta_3-\Delta_2}|x_2-x_3|^{\Delta_2+\Delta_3-\Delta_1}} \,,
\eeq
where the three-point function coefficient is,\footnote{Although we only consider the case of (2+1) dimensions, the three-point function coefficient can be easily generalized to that in $\mathbb{R}^{1,d+1}$:
\begin{align}
C(\Delta_1,\Delta_2;\Delta_3)=g\frac{m^{\Delta_1+\Delta_2-d-2}}{2^{\Delta_1+\Delta_2}}
\frac{\Gamma\(\frac{\Delta_1+\Delta_3-\Delta_2}{2}\)\Gamma\(\frac{\Delta_2+\Delta_3-\Delta_1}{2}\)}{\Gamma(\Delta_3)}\,.
\end{align}}
\beq\label{3pt}
C(\Delta_1,\Delta_2;\Delta_3)=g\frac{m^{\Delta_1+\Delta_2-3}}{2^{\Delta_1+\Delta_2}}
\frac{\Gamma\(\frac{\Delta_1+\Delta_3-\Delta_2}{2}\)\Gamma\(\frac{\Delta_2+\Delta_3-\Delta_1}{2}\)}{\Gamma(\Delta_3)}\,.
\eeq
Recall that $\Delta_3$ is the conformal dimension we assign to the massive particle. 

\subsection{Four-Point Function}

Let us now move on to a general discussion of four-point amplitudes written in the massless conformal basis in $\mathbb{R}^{1,2}$. 
Consider a massless scalar two-to-two scattering amplitude 
\begin{align}
\mathcal{T}(s,t)\delta^{(3)} (\sum_i \epsilon_i \omega_i q^\mu(x_i))
\end{align}
 in (2+1) dimensions. Here we  parametrize the null momenta $k^\mu_i = \epsilon_i\omega_i q^\mu(x_i)$ as in \eqref{qmap}  and $\epsilon_i=\pm1$ for an outgoing/incoming particle. $s,t,u$ are the Mandelstam variables defined as $s=- (k_1+k_2)^2,t=-(k_1+k_3)^2,u=-(k_1+k_4)^2$.   Constrained by the massless kinematics, nontrivial scattering process only exists if two of the $\epsilon_i$'s have the opposite signs than the other two.  Depending on which two of the particles are incoming and which two are outgoing, we have six different crossing channels for the two-to-two scattering process.  Using CPT, the six crossing channels reduce to three, which will be denoted as $12\leftrightarrow 34$, $13\leftrightarrow 24$, and $14\leftrightarrow 23$.  The Mandelstam variables $s,t,u$ have fixed signs in a given crossing channel.  For example, $s>0$ and $t,u<0$ in the $12\leftrightarrow 34$ channel.

Importantly, the amplitudes in the conformal basis depend on the choice of the crossing channels.  We will specify the crossing channel under consideration in the following discussion.  The crossing relations between these amplitudes will be discussed in Section \ref{sec:crossing}.


  In the massless conformal basis, the amplitude takes the form
\begin{align}
\widetilde{\mathcal{A}}(\Delta_i , x_i ) = \prod_{i=1}^4\int_0^\infty d\omega_i \omega_i^{\Delta_i-1} \, \mathcal{T}(s,t)\delta^{(3)} (\sum_i \epsilon_i \omega_i q_i^\mu(x_i))
\,.
\end{align}
Three of the four integrals can be done by solving the delta functions:
\begin{align}
\omega_2 =- \epsilon_1 \epsilon_2 {x_{13} x_{14} \over x_{23}x_{24}} \omega_1\,,~~~~~
\omega_3 = \epsilon_1 \epsilon_3 {x_{12} x_{14} \over x_{23} x_{34}} \omega_1\,,~~~~~
\omega_4 = -\epsilon_1\epsilon_4 {x_{12} x_{13} \over x_{24} x_{34}}\omega_1\,,
\end{align}
where $x_{ij}=x_i-x_j$. On the support of the delta function, the Mandelstam variables are 
\begin{align}
\begin{split}
&s=  -4 {x_{12}^2x_{13}x_{14}\over x_{23}x_{24}} \omega_1^2\,,~~~~
t = -  {1\over  z } s\,,~~~~
u  ={1- z  \over  z } s\,,
\end{split}
\end{align}
where the real cross ratio is
\begin{align}
 z  = {x_{12} x_{34} \over x_{13}x_{24}} \in\mathbb{R}\,.
\end{align}
The delta functions only have support when all the $\omega_i$'s are positive.  This constrains the real cross-ratio $ z $ in the following way
\begin{align}
\begin{split}
&12 \leftrightarrow 34\text{ channel}:~ z \in(1,\infty)\,,\\
&13\leftrightarrow 24\text{ channel}:~ z \in (0,1)\,,\\
&14\leftrightarrow 23\text{ channel} :~  z \in (-\infty,0)\,.
\end{split}
\end{align}
Indeed, for example in the $12\leftrightarrow 34$ channel, the scattering angle $\theta$ is related to the cross ratio $ z $ as $ z ^{-1} =  \sin^2(\theta/2)< 1$.

Let  $\epsilon_s\equiv\epsilon_1\epsilon_2 $, which is $+1$ in the $12\leftrightarrow 34$ channel and $-1$ in the other two channels.
Define $\omega$ as
\begin{align}
\omega \equiv \sqrt{ \epsilon_s s} =\sqrt{|s|}\,.
\end{align} 
 We then obtain
\begin{align}
&\widetilde{\mathcal{A}}(\Delta_i , x_i )
= \frac{2^{- \sum_i\Delta_i +1 }}{ 
|x_{23} x_{24} x_{34}|}\left| {x_{13} x_{14} \over x_{23} x_{24} }\right|^{\Delta_2-1}
\left| {x_{12} x_{14} \over x_{23} x_{34} }\right|^{\Delta_3-1}
\left| {x_{12} x_{13} \over x_{24} x_{34} }\right|^{\Delta_4-1}
\left| {x_{23} x_{24} \over x_{12}^2 x_{13}x_{14} } \right|^{\frac 12 \sum_i \Delta_i -\frac 32}\notag\\
&\times
\int_0^\infty d\omega \, \omega^{\sum_i\Delta_i -4}\, 
\mathcal{T}\left(\epsilon_s \omega^2 , -\epsilon_s\frac 1 z \omega^2 \right)\,.
\end{align}
For simplicity, let us consider the case when all four conformal  dimensions are the same $\Delta_i =\Delta_\phi$.  Then we have
\begin{align}\label{4ptmain}
&\widetilde{\mathcal{A}}(\Delta_\phi , x_i )
= {1\over |x_{12}|^{2\Delta_\phi} |x_{34}|^{2\Delta_\phi}} \,
f( z )\,,
\end{align}
where $ f( z )$ is
\begin{align}\label{main4pt}
\, f( z )
=
2^{-4\Delta_\phi+1}
{ | z | \over \sqrt{| z -1|} }
\int_0^\infty d\omega \, \omega^{4\Delta_\phi -4}\,  
\mathcal{T}\left(\epsilon_s \omega^2 , -\epsilon_s\frac 1 z  \omega^2 \right)\,\,.
\end{align}
Recall that $\epsilon_s=+1$ in the $12\leftrightarrow 34$ channel  and $\epsilon_s=-1$ in the $13\leftrightarrow 24$ and $14\leftrightarrow 23$ channels.

\begin{figure}[h!]
\begin{center}
\includegraphics[width=.9\textwidth]{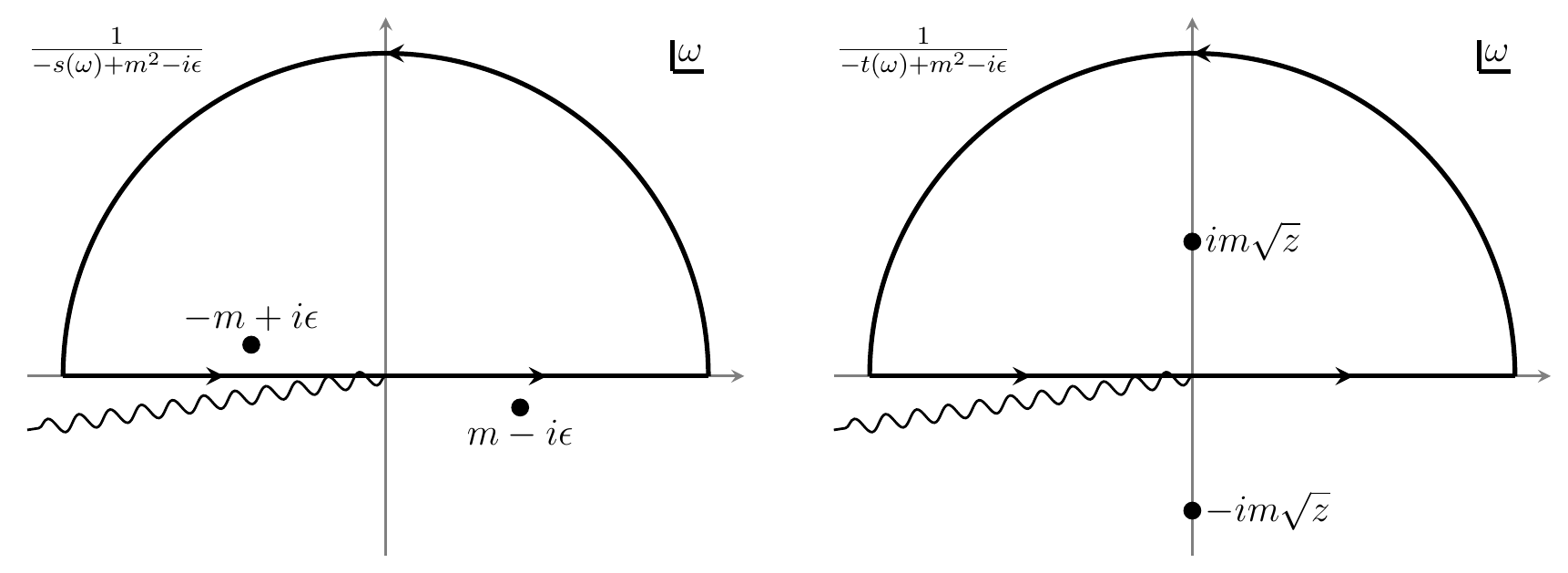}
\end{center}
\caption{The analytic structure of the integral \eqref{realline} for the tree-level exchange four-point function $f_{12\leftrightarrow 34}(z)$. Left: the $s$-channel contribution. Right: the $t$-channel contribution. }\label{fig:contour}
\end{figure}

Let us present an alternative formula in the case of tree-level scattering amplitudes. We will focus on  the $12\leftrightarrow 34$ channel  but the discussion can be easily generalized to other channels. The four-point function in this channel is
\begin{align}
 f_{12\leftrightarrow 34}( z )
=2^{-4\Delta_\phi+1}
{  z  \over \sqrt{ z -1} }
\int_0^\infty d\omega \, \omega^{ 4\Delta_\phi -4 }\, \mathcal{T}\left(\omega^2 , -\frac 1 z  \omega^2 \right)\,,~~~~ z >1
\end{align}
Let us discuss the analytic property of this integral.  For tree-level amplitudes,  $\mathcal{T}$ only  has poles in $\omega^2$. In addition, the integrand has a branch cut emitting from $\omega=0$.  We can choose the branch cut to be almost aligned with the negative real line, but slightly below it. With this choice of the branch cut, we can extend the integral to be over the full real line
\begin{align}\label{realline}
& f_{12\leftrightarrow 34}( z )
=  {2^{-4\Delta_\phi+1} \over 1+e^{4\pi i \Delta_\phi}}
{  z  \over \sqrt{ z -1} }
\int_{-\infty}^\infty d\omega \, \omega^{ 4\Delta_\phi -4 }\,  
\mathcal{T}\left(\omega^2 , -\frac 1 z  \omega^2 \right)\,,~~~ z >1\,.
\end{align}
See Figure \ref{fig:contour} for the example of exchange diagrams \eqref{exchange}. 
If we further assume the following fall-off condition on the upper half plane of $\omega$,
\begin{align}\label{falloff}
\lim _{|\omega|\to \infty} \omega^{4\Delta_\phi-3}
\mathcal{T}\left(\omega^2 , -\frac 1 z  \omega^2 \right) \to0\,,
\end{align}
and that there is  no singularity at $\omega=0$, then
 the contour  can be closed from above.  We obtain
\begin{align}\label{residue}
{f}^{tree}_{12\leftrightarrow 34}( z )
=  2^{-4\Delta_\phi+1}{2\pi i  \over 1+e^{4\pi i \Delta_\phi}}
{  z  \over \sqrt{ z -1} }
\sum_{\omega^*}\text{Res}_{\omega=\omega^*}
 \, \omega^{ 4\Delta_\phi -4 }\,  \mathcal{T}\left(\omega^2 , -\frac 1 z  \omega^2 \right)\,,~~~ z >1\,,
\end{align}
where the sum is over all poles on the upper half $\omega$-plane.

\subsection{Crossing Symmetry}\label{sec:crossing}

In \eqref{4ptmain} we have presented a general formula for the two-to-two massless scalar amplitude in the conformal basis.  The resulting correlator depends explicitly on the crossing channel, i.e. it depends on which particles we take to be outgoing and incoming.  Thus given a single amplitude in momentum space $\mathcal{T}(s,t)$, we end up with three correlators in the conformal basis $ f_{12\leftrightarrow 34}( z ),f_{13\leftrightarrow 24}( z ), f_{14\leftrightarrow23}( z )$.

Let us study what  crossing symmetry in momentum space implies on these three four-point functions in the conformal basis.  
We will consider scattering amplitudes of identical particles and assume the  amplitude in momentum space has the $s\leftrightarrow t\leftrightarrow u$ crossing symmetry,
\begin{align}
\mathcal{T}(s,t) =\mathcal{T}(t,s) = \mathcal{T}(-s-t,t)\,.
\end{align}
The amplitude crossing symmetry implies that of the correlator in the conformal basis.  Indeed, 
\begin{align}
 f_{13\leftrightarrow24} (1- z ) &=  2^{-4\Delta_\phi+1}
{1- z  \over \sqrt{ z } } \int_0^\infty d\omega \omega^{4\Delta_\phi-4}
\mathcal{T}\left( -{ z \over 1- z } \omega^2 , {1\over 1- z  }\omega^2\right)\notag\\
&= 2^{-4\Delta_\phi+1} \left( {1- z \over  z }\right)^{2\Delta_\phi} { z \over \sqrt{1- z }} 
\int_0^\infty d\omega' \omega'^{4\Delta_\phi-4} \mathcal{T}\left(-\omega'^2 , \frac 1 z  \omega'^2\right)\,,
\end{align}
where in the first line we have used $\mathcal{T}(s,t)=\mathcal{T}(-s-t,t)$, and in the second line we have  rescaled $\omega' = \sqrt{  z \over 1- z  } \omega$.  Hence we conclude that $f_{13\leftrightarrow 24}$ satisfies the crossing equation:
\begin{align}\label{1324crossing}
 f_{13\leftrightarrow24} ( 1- z   ) = \(\frac{1-z}{z}\)^{2\Delta_\phi}f_{13\leftrightarrow 24} (  z )\,,~~~~~0< z  <1\,.
\end{align}
Using a similar argument, crossing symmetry also relates the three correlators from different crossing channels:
\begin{align}
 f_{13\leftrightarrow 24} ( z )
= { z ^{2\Delta_\phi}}  f_{12\leftrightarrow 34} (1/ z )
=   f_{14\leftrightarrow 23} \left( { z \over z-1 }\right)\,,~~~~~0< z <1\,.
\end{align}
 There is one caveat regarding the relation between $f_{14\leftrightarrow 23}$ and $ f_{13\leftrightarrow24}$.  In both the $14\leftrightarrow 23$ and $13\leftrightarrow 24$ channels, we have $\epsilon_s=-1$ in \eqref{4ptmain}. 
One might then naively equate $ f_{13\leftrightarrow 24}( z )$ with  $-  f_{14\leftrightarrow 23}( z )$, with the latter extended to $0< z <1$. 
This is generally ambiguous because the latter was originally defined only for $ z <0$ in \eqref{4ptmain}, and the extension to $0< z <1$ requires a choice of the $i\epsilon$ prescription.  More explicitly, the $i\epsilon$ prescription should be such that both $t = \frac 1 z  \omega^2$ and $u={ z -1\over  z } \omega^2$ get shifted by $+i\epsilon$, which is the standard $i\epsilon$ prescription in momentum space.  
However, one can easily see that there is no such $i\epsilon$ prescription for $ z $ such that $t( z )\to t( z )+i\epsilon$ \textit{and} $u( z )\to u( z )+i\epsilon$.\footnote{For example, in the $14\leftrightarrow 23$ channel where $ z <0$,  if one chooses to continue $ z \to  z -i\epsilon$, then $t( z )\to t( z )+i\epsilon $ but $u( z )\to u( z ) -i\epsilon$.}  In other words,  $f_{13\leftrightarrow 24}( z )$ with  $-  f_{14\leftrightarrow 23}( z )$ are not related by an analytic continuation in the real cross ratio $ z $. One can verify this explicitly in the examples of tree-level exchange amplitude in \eqref{phi3t} and \eqref{phi3u}  below.

\subsection{Tree-Level Exchange Amplitude}\label{sec:scalarexchange}

Let us return to the scalar theory considered in Section \ref{sec:3pt}.  
The tree-level amplitude with massless external scalar particles exchanging a massive scalar is
\begin{align}\label{exchange}
\mathcal{T}(s,t) = {g^2\over -s+m^2 -i\epsilon}
+{g^2\over -t+m^2 -i\epsilon}
+{g^2\over -u+m^2 -i\epsilon}\,.
\end{align}
The massive scalar exchange amplitude is in some sense simpler than  other  tree-level amplitudes, for example the contact four-point or the massless exchange amplitudes. Indeed, the latter two suffer from either the UV or IR divergence for positive $\Delta_\phi$ in the change of basis integral \eqref{main4pt}. On the other hand, for the massive exchange amplitude,  the mass $m$ of the intermediate particle provides an IR cut-off so that there is range of $\Delta_\phi$ where the amplitude in the conformal basis is well-defined.  In fact, for sufficiently positive $\Delta_\phi$, there is a good physical reason for the divergence of the contact four-point and the massless exchange amplitudes in the conformal basis: they   are exactly the bulk point singularity in $AdS_3/CFT_2$ of the corresponding  Witten diagrams. We will come back to this point in Section \ref{sec:bulkpoint}.

Let us start with the $12\leftrightarrow 34$ channel where we take particles 1 and 2 to be incoming while 3 and 4 to be outgoing.  Assuming $3/4 <\text{Re} \,\Delta_\phi <5/4$,  this amplitude \eqref{exchange} satisfies the fall-off condition \eqref{falloff}, we can directly apply the residue formula \eqref{residue} to obtain the amplitude in the $12\leftrightarrow 34$ channel:
\begin{align}\label{phi3s}
 f_{12\leftrightarrow 34}( z ) = 
 \mathcal{N}_{\Delta_\phi}\,
{ z \over \sqrt{ z -1}}
\left[ \,
e^{i \pi \alpha}
+ z ^{\alpha}
+\left( { z \over  z -1}\right)^{\alpha}
\, \right]\,,~~~ z >1\,,
\end{align}
where $\mathcal{N}_{\Delta_\phi}=g^2 { \pi m^{4\Delta_\phi-5}\over2^{4\Delta_\phi}\cos (2\pi \Delta_\phi) }$ and
\begin{align}
\alpha\equiv 2\Delta_\phi -\frac32\,.
\end{align}
Similarly, in the $13\leftrightarrow 24$ channel, the four-point function is given by
\begin{align}\label{phi3t}
 f_{13\leftrightarrow 24}( z ) =  \mathcal{N}_{\Delta_\phi}\,
{ z \over \sqrt{1- z }}
\left[ \,
1
+e^{i \pi \alpha} z ^{\alpha}
+\left( { z \over 1- z }\right)^{\alpha}
\, \right]\,,~~0< z <1\,.
\end{align}
Finally in the $14\leftrightarrow 23$ channel, the four-point function is given by
\begin{align}\label{phi3u}
f_{14\leftrightarrow 23}( z ) =  \mathcal{N}_{\Delta_\phi}\,
{(- z )\over \sqrt{1- z }}
\left[ \,
1
+(- z )^{\alpha}
+e^{i \pi\alpha}\left( {- z \over 1- z }\right)^{\alpha}
\, \right]\,,~~ z <0\,.
\end{align}
Even though the change of basis integral only converges for $3/4<\text{Re} \, \Delta_\phi<5/4$, we can analytically continue the final expression to all complex $\Delta_\phi$ except for $\Delta_\phi \in \frac14 +\mathbb{Z}$. 

One feature of these correlators is that they are complex  even if we assume $\Delta_\phi$ to be real.  This is of course expected because the amplitude is already complex in momentum space because of  the $i\epsilon$ prescription.  It follows that the imaginary part of these correlators should obey the optical theorem in the conformal basis, which we will discuss in Section \ref{sec:optical}.

\section{Optical Theorem and Conformal Block Decomposition}\label{sec:optical}

In this section we translate the optical theorem into the conformal basis.  In the case of the tree-level exchange amplitude, the optical theorem takes the form of a conformal block decomposition on the principal continuous series, with  coefficients being the three-point amplitude in the conformal basis.  We then verify this explicitly in a scalar theory in (2+1) dimensions. 

\subsection{Conformal Optical Theorem}

Let us start with the simplest example of a tree-level massive scalar exchange amplitude $\mathcal{T}(s,t)$ in \eqref{exchange} in $\mathbb{R}^{1,d+1}$. The optical theorem relates the imaginary part of the four-point amplitudes to the product of two three-point amplitudes
\beq
\label{optical2}
2\,\text{Im}\,\mathcal{A}_{k_1,k_2\rightarrow k_3,k_4}=
\pi
\int { d^{d+1}\vec p\over \sqrt{\vec p^2+m^2}}
\mathcal{A}_{k_1,k_2\rightarrow p}\mathcal{A}_{k_3,k_4 \rightarrow p}\,.
\eeq
where $\mathcal{A}_{k_1,k_2\rightarrow k_3,k_4} =\mathcal{T}(s,t) \delta^{(d+2)}(k_1+k_2-k_3-k_4)$ and $\mathcal{A}_{k_1,k_2\rightarrow p}= g \delta^{(d+2)} (k_1+k_2- p)$. Note that this simplest optical theorem  follows from the identity:
\beq
\text{Im}\frac{1}{-s+m^2-i\epsilon}=\pi \delta(s-m^2)\,.
\eeq
Since we take particles 1 and 2 to be incoming while 3 and 4 to be outgoing, only the $s$-channel can go on-shell and contribute to the imaginary part of the amplitude. 


To go to the conformal basis, we use the following orthogonality condition of the AdS bulk-to-boundary propagator \cite{Costa:2014kfa} (see also \cite{Pasterski:2017kqt}):
\beq\label{otho}
\int^{\infty}_{-\infty} d \nu\mu(\nu)\int d^d \vec x\, G_{\frac{d}{2}+i\nu}(\hat{p}_1;\vec x)G_{\frac{d}{2}-i\nu}(\hat{p}_2;\vec x)=(\hat{p}_1)^0\,\delta^{(d+1)}(\hat{p}_1-\hat{p}_2)\,,
\eeq
where $\hat p_i$ are unit timelike vectors in $\mathbb{R}^{1,d+1}$ and the right-hand side is the $SO(1,d+1)$ invariant delta-function. The measure $\mu(\nu)$ is
\beq\label{measure}
\mu(\nu)=\frac{\Gamma\(\frac{d}{2}+i\nu\)\Gamma\(\frac{d}{2}-i\nu\)}{4\pi^{d+1}\Gamma(i\nu)\Gamma(-i\nu)}\,.
\eeq
We can then insert $1=\int {d^{d+1} \hat\ell  \over \hat\ell^0 }\hat{\ell}^0\,\delta^{(d+1)}(\hat{\ell}-\hat{p})$  into the right-hand side of \eqref{optical2} and use \eqref{otho} to obtain
\beq
&2\,\text{Im}\, \mathcal{A}_{k_1,k_2\rightarrow k_3,k_4}
\\
=&\pi m^d\int^{\infty}_{-\infty} d  \nu\mu(\nu)\int d^d \vec x\,\int\frac{d^{d+1}\hat{p}}{\hat{p}^0}\mathcal{A}_{k_1,k_2\rightarrow p}G_{\frac{d}{2}+i\nu}(\hat{p};\vec x)\int\frac{d^{d+1} \hat{\ell}}{\hat{\ell}^0}\mathcal{A}_{k_3,k_4 \rightarrow \ell}G_{\frac{d}{2}-i\nu}(\hat{\ell};\vec x)\,.
\eeq
Now we perform the change of basis integral \eqref {changebasis} on the external particles $k_1,\cdots, k_4$.  In the conformal basis, the optical theorem is then translated into\footnote{We assume the conformal dimensions $\Delta_i$ of the external operators can be analytically continued to be real.}
\begin{align}
&2\, \text{Im}\, \widetilde{\mathcal{A}}_{4}(\Delta_i ,\vec x_i)\\
&
=\pi m^d\int^{\infty}_{-\infty}d \nu\mu(\nu)\int d^d \vec x\, 
\widetilde{\mathcal{A}}_{3}\left( \Delta_1, \Delta_2, {d\over2}+i\nu ;
\vec x_1,\vec x_2,\vec x\right)
\widetilde{\mathcal{A}}_{3}\left( {d\over2}-i\nu,\Delta_3, \Delta_4,  ;
\vec x,\vec x_3,\vec x_4\right)\,,\notag
\end{align}
where we have assigned conformal dimension $\Delta_i$ to the $i$-th external particle. 
 The  optical theorem can be further simplified by using the explicit positions dependence of the three-point functions:
 \begin{align}\label{3ptagain}
 \widetilde{\mathcal{A}}_3 (\Delta_i ,\vec x_i )  ={ C(\Delta_1,\Delta_2 ;\Delta_3) \over
 |\vec x_{12}|^{\Delta_1+\Delta_2-\Delta_3}  |\vec x_{13}|^{\Delta_1 +\Delta_3 -\Delta_2} 
 |\vec x_{23}|^{\Delta_2 +\Delta_3-\Delta_1}
  } \,.
 \end{align}
  The integral over $\vec x$ gives the shadow representation of the conformal partial wave $\Psi_\Delta(z,\bar z)$ \cite{Ferrara:1972uq,Dolan:2000ut,Dolan:2011dv,SimmonsDuffin:2012uy}:
\beq
&\frac{\Psi_\Delta(z,\bar{z})}{|\vec x_{12}|^{\Delta_1+\Delta_2}|\vec x_{34}|^{\Delta_3+\Delta_4}}\left|\frac{\vec x_{24}}{\vec x_{14}}\right|^{\Delta_{12}}\left|\frac{\vec x_{14}}{\vec x_{13}}\right|^{\Delta_{34}}
\\
=&\frac{1}{2}\int d ^d \vec w\frac{|\vec x_{12}|^{\Delta-\Delta_1-\Delta_2}|\vec x_{34}|^{d-\Delta-\Delta_3-\Delta_4}}{|\vec w-\vec x_{1}|^{\Delta_{12}+\Delta}|\vec w-\vec x_{2}|^{\Delta_{21}+\Delta}|\vec w-\vec x_{3}|^{\Delta_{34}+d-\Delta}|\vec w-\vec x_{4}|^{\Delta_{43}+d-\Delta}}\,,
\eeq
where $\vec x_{ij}= \vec x_i -\vec x_j$ and $\Delta_{ij} =\Delta_i - \Delta_j$. $z,\bar z$ are the cross ratios:
\begin{equation}
\begin{gathered}
z\bar{z}=\frac{|\vec x_{12}|^2|\vec x_{34}|^2}{|\vec x_{13}|^2|\vec x_{24}|^2}\,,
\quad (1-z)(1-\bar{z})=\frac{|\vec x_{14}|^2|\vec x_{23}|^2}{|\vec x_{13}|^2|\vec x_{24}|^2}\,.
\end{gathered}
\end{equation}
$\Psi_\Delta(z,\bar{z})$ is an eigenfunction of the conformal Casimir operator  that is shadow symmetric, i.e. $\Psi_\Delta = \Psi_{ d-\Delta}$. It is related to the scalar conformal blocks $G_{\Delta}^{(\ell=0)}(z,\bar{z})$ with intermediate scalar primaries as \cite{Dolan:2011dv},
\beq
\Psi_\Delta(z,\bar{z})=\frac{c_\Delta G_{\Delta}^{(\ell=0)}(z,\bar{z})+c_{d-\Delta}G^{(\ell=0)}_{d-\Delta}(z,\bar{z})}{
2\Gamma\(\frac{\Delta+\Delta_{21}}{2}\)\Gamma\(\frac{\Delta-\Delta_{21}}{2}\)
\Gamma\(\frac{d-\Delta+\Delta_{34}}{2}\)\Gamma\(\frac{d-\Delta-\Delta_{34}}{2}\)
}\,,
\eeq
where the coefficient $c_\Delta$ is
\beq
c_\Delta=\pi^{d/2}\frac{\Gamma\(\frac{d}{2}-\Delta\)\Gamma\(\frac{\Delta+\Delta_{34}}{2}\)\Gamma\(\frac{\Delta-\Delta_{34}}{2}\)\Gamma\(\frac{\Delta+\Delta_{21}}{2}\)\Gamma\(\frac{\Delta-\Delta_{21}}{2}\)}{\Gamma\(\Delta\)}\,.
\eeq
If we write the four-point function as
\begin{align}
\widetilde{\mathcal{A}}_4(\Delta_i ,\vec x_i )  = {|\vec x_{12}|^{\Delta_1+\Delta_2}|\vec x_{34}|^{\Delta_3+\Delta_4}}\left|\frac{\vec x_{24}}{\vec x_{14}}\right|^{\Delta_{12}}\left|\frac{\vec x_{14}}{\vec x_{13}}\right|^{\Delta_{34}} f(z,\bar z)\,,
\end{align}
then the optical theorem gives  the conformal block decomposition for the imaginary part of $f(z,\bar z)$ on the principal continuous series
\beq\label{CBD}
\boxed{\,
\text{Im}f(z,\bar{z})=\pi m^d\int^{\infty}_{-\infty} d \nu\mu(\nu)C\(\Delta_1,\Delta_2;\frac{d}{2}+i\nu\)C\(\Delta_3,\Delta_4;\frac{d}{2}-i\nu\)\Psi_{\frac{d}{2}+i\nu}(z,\bar{z})\,,}
\eeq
with coefficient being the three-point amplitude written in the conformal basis \eqref{3ptagain}.

The derivation can be extended to the general optical theorem straightforwardly,
\beq
\label{optical}
2\text{Im}\mathcal{T}_{\{p\}\rightarrow \{q\}}=\sum_{\{k\}} \prod_{j}\(\int\frac{ d ^{d+1} \vec{k}_j}{(2\pi)^{d+1} 2 k^0_j}\)\mathcal{T}_{\{p\}\rightarrow \{k\}}\mathcal{T}_{\{q\} \rightarrow \{k\}}(2\pi)^{d+2}\delta^{(d+2)}\(\sum p-\sum k\)\,,
\eeq
where $p,q,k$ are the sets of momenta of the initial, final and intermediate particles. The sum in $\{k\}$ is over all possible intermediate particle states. 
The general optical theorem when translated into the conformal basis is\footnote{Here for simplicity we have assumed all the intermediate particles are massive scalars with masses $m_j$.   
A similar formula holds true for massless intermediate scalars with the measure factor $m_j^d \mu(\nu_j)$ replaced by $2^d$.  The apparent mass dimension mismatch comes from our normalization of the massless versus massive conformal primary wavefunctions.
}
\begin{align}
2\text{Im}\widetilde{\mathcal{A}} (\Delta_I ,\Delta_F, \vec x_i) = &(2\pi)^{d+2}\sum_{\{k\}}\prod_{j}\(\frac{m_j^d}{2(2\pi)^{d+1}}\int^{\infty}_{-\infty}d \nu_j\mu(\nu_j)\int d^d \vec w_j\) \\
&~~~~~~~~~~\times\widetilde{\mathcal{A}}\left(\Delta_I ,{d\over 2}+i\nu_j ;\vec x_I ,\vec w_j\right)
\widetilde{\mathcal{A}}\left( {d\over 2}-i\nu _j , \Delta_F ; \vec w_j ,\vec x_F\right)\,.\notag
\end{align}
where $I$ and $F$ are the subscripts for  the conformal dimensions and the positions of the initial and final conformal primary wave functions, respectively.

\subsection{Example: Tree-Level Exchange Amplitude}
We now explicitly show that, in the case of the exchange diagram in (2+1) dimensions,  the conformal block decomposition of the four-point function in \eqref{phi3s} reproduces  the three-point function coefficient in \eqref{3pt}. In one dimension, the conformal partial wave $\Psi_{\frac{d}{2}+i\nu}(z)$, defined for $z>0$, has been worked out in \cite{Maldacena:2016hyu} (see also \cite{Hogervorst:2017sfd})
\beq
\label{MSbasis}
\Psi_\Delta(z)=
\begin{dcases}
\frac{\tan\pi \Delta}{2\tan \frac{\pi \Delta}{2}}\frac{\Gamma(\Delta)^2}{\Gamma(2\Delta)}z^\Delta {}_2F_1(\Delta,\Delta;2\Delta;z)+(\Delta\leftrightarrow 1-\Delta),\quad &0<z<1\,,
\\
{\Gamma(\frac 12 -\frac \Delta2) \Gamma(\frac \Delta2) \over \sqrt{\pi}}
\,_2F_1\left( \frac \Delta2 , \frac12 - \frac \Delta2 ;\frac 12;{(2-z)^2\over z^2 }\right),\quad &1<z<2\,,
\\
\Psi_\Delta\(\frac{z}{z-1}\),\quad & 2<z\,.
\end{dcases}
\eeq
$\Psi_\Delta(z)$'s on the principal discrete series $\Delta\in 2\mathbb{N}$ together with the principal continuous series $\Delta\in\frac{1}{2}+i\mathbb{R}_{\ge0}$  form an orthogonal basis for the space of functions with the following boundary conditions (1) 
$f(z)=f(\frac{z}{z-1})$ for $z\in(1,\infty)$ which in particular implies  $f'(2)=0$, and (2) $f(z)$ vanishes no slower than $z^{1/2}$ as $z\rightarrow0$. 
The inner product on this space of functions is
\beq
\langle g,f\rangle=\int^2_0\frac{ d  z}{z^2}g^*(z)f(z)=\(\int^1_0+\frac{1}{2}\int_1^\infty\)\frac{ d  z}{z^2}g^*(z)f(z)\,,
\eeq
where the integral between  $(1,2)$ is replaced by an integral between $(1,\infty)$ using the symmetry $f(z)=f(\frac{z}{z-1})$. The inner products of $\Psi_\Delta$ are
\beq
\langle \Psi_{\Delta},\Psi_{\Delta'}\rangle&=\frac{\pi^2}{4\Delta-2}\delta_{\Delta,\Delta'}, ~~~~~~~~~~~~~~~~~~\,\Delta,\Delta'\in2\mathbb{N}
\\
\langle \Psi_{\Delta},\Psi_{\Delta'}\rangle&=\frac{\pi^2\tan\pi \Delta}{2\Delta-1}\delta(\Delta-\Delta'),\quad~~~~~ \Delta,\Delta'\in\frac 12+ i\mathbb{R}_{\ge0}\,.\,
\eeq

Let us decompose the four-point function $f_{12\leftrightarrow34}(z)$ in \eqref{phi3s} on this basis.   The  imaginary part of $ f_{12\leftrightarrow 34}$ can be written as
\beq
2\text{Im}{f}_{12\leftrightarrow34}(z)=2\mathcal{N}_{\Delta_\phi}\sin\(2\pi\Delta_\phi-\frac{3\pi}{2}\)F(z),\quad F(z)=
\begin{dcases}
\frac{z}{\sqrt{z-1}}, &1<z
\\
0,&0<z<1\,.
\end{dcases}
\eeq
Recall that the real cross ratio $z$ in $12\leftrightarrow 34$ channel is constrained by the flat space kinematics to be $z>1$. 
This function satisfies the two boundary conditions: $F(z)=F(\frac{z}{z-1})$ for $z\in(1,\infty)$ and $\lim_{z\rightarrow0}F'(z)/z^{1/2}=0$. Thus it can be expanded on the basis with the coefficients proportional to the inner product
\beq
\langle F,\Psi_\Delta\rangle=\frac{1}{2}\int^\infty_1\frac{ d  z}{z^2}\frac{z}{\sqrt{z-1}}\Psi_\Delta(z)=\frac{\pi^3}{\sin(\pi \Delta)\Gamma(\frac{2-\Delta}{2})^2\Gamma(\frac{\Delta+1}{2})^2}\,.
\eeq
Since the inner products vanishes on the principal discrete series $h\in2\mathbb{N}$, $F(z)$ can be expanded just on the principal continuous series,
\beq
F(z)=\int^{\frac{1}{2}+i\infty}_{\frac{1}{2}-i\infty}\frac{ d  \Delta}{2\pi i}  
\frac{2\Delta-1}{\pi\tan(\pi \Delta)}\frac{\pi^3}{\sin(\pi \Delta)\Gamma(\frac{2-\Delta}{2})^2\Gamma(\frac{\Delta+1}{2})^2}\Psi_\Delta(z)\,,
\eeq
where we have used the symmetry  $\Delta\leftrightarrow 1-\Delta$  of the integrand to extend the integration from $\frac 12+i\mathbb{R}_{\ge0}$ to $\frac 12+i\mathbb{R}$. 
The coefficient of the above decomposition can be written in terms of the three-point function coefficient \eqref{3pt} as
\beq
\text{Im}f_{12\leftrightarrow34}(z)&
&=\pi m\int^{\infty}_{-\infty} d \nu\mu(\nu)C\(\Delta_1,\Delta_2;\frac{d}{2}+i\nu\)C\(\Delta_1,\Delta_2;\frac{d}{2}-i\nu\)\Psi_{\frac{d}{2}+i\nu}(z,\bar{z})\,,
\eeq
where  $C(\Delta_\phi,\Delta_\phi;\Delta)=g\frac{ m^{2\Delta_\phi-3}}{2^{2\Delta_\phi}}\frac{\Gamma\(\Delta/2\)^2}{\Gamma(\Delta)}$.  
Hence we have checked the conformal block decomposition \eqref{CBD} in the (2+1)-dimensional scalar theory with cubic coupling.

\section{The Bulk Point Singularity in AdS/CFT}\label{sec:bulkpoint}

In this section we discuss the relation between the massless conformal basis in $\mathbb{R}^{1,d+1}$ and the bulk point singularity in  $AdS_{d+2}/CFT_{d+1}$ in  Lorentzian signature.  In particular we argue that  scalar exchange amplitudes in the conformal basis discussed in Section \ref{sec:scalarexchange} arise from approaching the bulk point singularity at the same time scaling the intermediate conformal dimension to infinite in the exchange Witten  diagram.

Let us begin with a general review on the bulk point singularity in $AdS/CFT$ and its relation to the flat space limit \cite{Polchinski:1999ry,Giddings:1999jq,Gary:2009ae,Heemskerk:2009pn,Penedones:2010ue,Okuda:2010ym,Maldacena:2015iua}.   
 Consider $AdS_{d+2}$ embedded in $\mathbb{R}^{2,d+1}$ as the locus $Y^I Y_I = -R^2$, where $R$ is the AdS radius.  Here we  use $Y^I$ with $I=-1,0,1,\cdots ,d+1$ as the flat coordinates of the embedding space $\mathbb{R}^{2,d+1}$ and the index $I$ is raised and lowered by the flat metric $\eta_{IJ}=\text{diag}(-1,-1,+1,\cdots, +1)$.   On the other hand, a point on the boundary of $AdS_{d+2}$ is represented by a null ray $P^I\sim \lambda P^I$ with $P^I P_I=0$. 


Consider an $n$-point Witten diagram with  boundary operators located at $P^I_a$, $a=1,\cdots, n$.   We will restrict ourselves to the case\footnote{This in particular applies to the four-point function in $AdS_3$, which we will pay special attention to later on.} $n=d+3$ so that in this case the vertex of the Landau diagram is only a point in $AdS_{d+2}$. 
The conditions for the bulk point singularity are that there exists a bulk point $Y_0$ such that
\begin{enumerate}
\item $Y_0$ is lightlike separated from all the boundary points $P_a$,
\begin{align}
\eta_{IJ} \, Y^I_0  P^J_a=0\,,~~~~\forall \,a=1,2,\cdots,n\,.\label{bulkpt1}
\end{align}
\item There exist $n$ ``frequencies" $\omega_a>0$ such that the momentum is conserved at $Y_0$:
\begin{align}
\sum_{a=1}^n \omega_a  P^I_a=0\,.\label{bulkpt2}
\end{align}
\end{enumerate}
In general such a bulk point $Y_0$ does not exist for generic boundary points $P_a$.  In other words, the bulk point singularity only arises when we place the boundary points at some specific configuration.  
For example, in the case of four-point functions, the second condition \eqref{bulkpt2} implies that the 4$\times4$ matrix $P_a \cdot P_b$ has a zero eigenvector $\omega_a$, and hence it has vanishing determinant. This in turns implies the cross ratios are real at the bulk point singularity configuration, i.e. $z=\bar z$.

\begin{figure}[h!]
\begin{center}
\includegraphics[width=.5\textwidth]{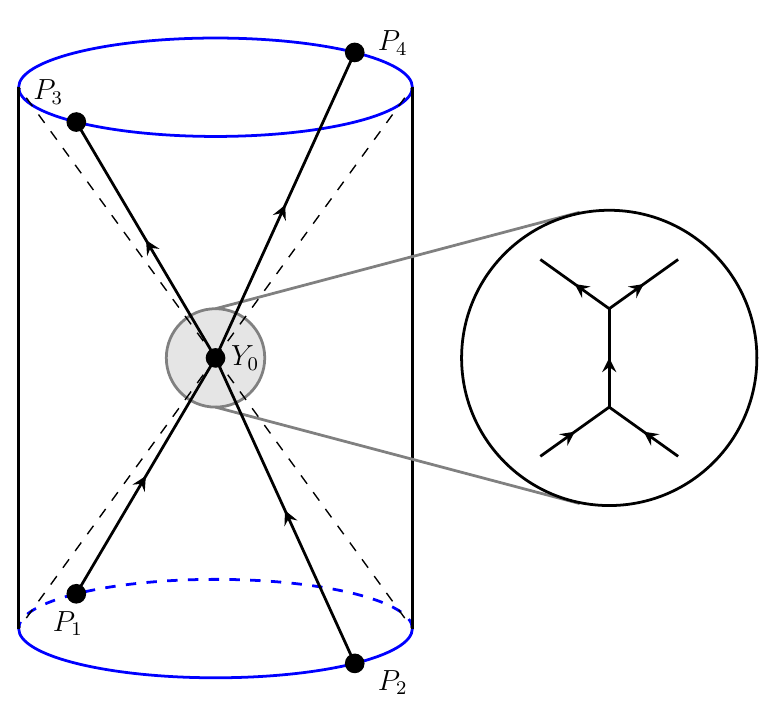}
\end{center}
\caption{The bulk point singularity in $AdS_3$. At the bulk point singularity configuration for a four-point function in $AdS_3/CFT_2$, the boundary points are restricted to two constant time slices in $AdS_3$, which are the future and past celestial circles (shown in blue) in the flat space limit to $\mathbb{R}^{1,2}$. The contribution to the Witten diagram is dominated by the integral around the bulk point $Y_0$, which can then be approximated by the flat space scattering process in the massless conformal basis.}\label{fig:bulkpoint}
\end{figure}

The bulk point singularity configuration is related to the flat space scattering kinematics in $\mathbb{R}^{1,d+1}$.   Let us take our  reference bulk point  to be at $Y_0 = (R,0,\cdots,0).$ 
Then the first condition \eqref{bulkpt1}  constrains the boundary points $P^I_a$ to be 
\begin{align}\label{Pq}
P^I_a  =  (0 , \pm q^\mu_a)\,,
\end{align}
where $q^\mu_a$ is  a null ray in $\mathbb{R}^{1,d+1}$ with $\mu = 0,1,\cdots,d+1$.  Here we take the time component of $q^\mu$ to be positive, so that the plus/minus sign above corresponds to a null vector in the future/past lightcone of $\mathbb{R}^{1,d+1}$ in the embedding space, respectively.  In other words, the boundary points  are restricted to two constant time slices in $AdS_{d+2}$ at the bulk point singularity configuration (see Figure \ref{fig:bulkpoint}).  
The null vectors $q^\mu_a$ are later identified as the directions of null momenta in the flat space scattering process.  
 Now the second condition \eqref{bulkpt2} states that there exists $n$ frequencies $\omega_a$ such that the flat space  momentum conservation $\sum_a \pm\omega_a q^\mu_a=0$ holds true.  In the case of the bulk point singularity for four-point functions, the real cross ratio $z=\bar z$ parametrizes the scattering angle. 

We now argue that perturbative scattering amplitudes in the massless conformal basis in $\mathbb{R}^{1,d+1}$ can be embedded into the $AdS_{d+2}$ Witten diagram with the same interaction at the bulk point singularity.  
The scalar bulk-to-boundary propagator in $AdS_{d+2}$ between  a bulk point $Y$ and a boundary point $P$ with conformal dimension $\Delta_\phi$ is
\begin{align}
G_{\Delta_\phi}(Y,P) = {\mathcal{C}_\Delta \over  R^{d\over 2}} {1\over (-2P\cdot Y/R + i \epsilon)^{\Delta_\phi}}\,,
\end{align}
where $\mathcal{C}_\Delta = \Gamma(\Delta) /(2\pi^{d+1\over2} \Gamma(\Delta-\frac {d-1}{2}))$.  
At the bulk point singularity configuration, from \eqref{Pq}, the null ray $P_a^I$ in $\mathbb{R}^{2,d+1}$ is restricted to  a null ray $q_a^\mu$ in $\mathbb{R}^{1,d+1}$. Furthermore, the contribution to the Witten diagram receives dominant contributions from around  $Y_0$, and we can approximate the bulk point integral in $Y$ by a flat space integral in $X^\mu \in\mathbb{R}^{1,d+1}$ around $Y_0$.  In this limit, the $AdS_{d+2}$ bulk-to-boundary propagator becomes proportional to the massless conformal primary wavefunction $\varphi^\pm_{\Delta_\phi}$ \eqref{masslessCPW} in $\mathbb{R}^{1,d+1}$:
\begin{align}
G_{\Delta_\phi}(Y,P ) \,\underset{\text{bulk point singularity}} { \longrightarrow}
\,\varphi_{\Delta_\phi}^\pm (X, q)\,.
\end{align}
  Whether the corresponding conformal primary wavefunction is incoming or outgoing is determined by which of the past and future time slices the boundary point $P^I_a$ is located at (i.e. the sign in \eqref{Pq}).


Following the same argument in \cite{Gary:2009ae,Maldacena:2015iua}, let us see how this works explicitly for the tree-level scalar exchange Witten diagram in $AdS_3/CFT_2$.  We will focus only on the $s$-channel diagram, while the other two follow identically.  

The scalar bulk-to-bulk propagator with intermediate conformal dimension $\Delta$ is (see, for example, \cite{Hijano:2015zsa})
\begin{align}
\Pi_\Delta (Y_1,Y_2) = {1\over 2\pi R} {e^{-\Delta\sigma }\over 1-e^{-2\sigma}}\,,
\end{align}
where $\sigma$ is the geodesic distance between $Y_1,Y_2$:
\begin{align}
\sigma(Y_1,Y_2) = \log\left(   {  1+\sqrt{1-\xi^2 +i\epsilon} \over \xi}\right)\,,~~~~~~
\xi  = - {R^2\over Y_1\cdot Y_2}\,.
\end{align}
The $s$-channel tree-level scalar Witten diagram with internal dimension $\Delta$ and  identical external operator dimension $\Delta_\phi$  is
\begin{align}\label{I4}
I_4&  =g^2 \int_{AdS_3} dY_1 \int_{AdS_3} dY_2\,
  G_{\Delta_\phi}(Y , P_1 )  G_{\Delta_\phi} (Y,P_2)  \Pi_\Delta(Y_1,Y_2)
  G_{\Delta_\phi}(Y,P_3 ) G_{\Delta_\phi}(Y,P_4)\notag\\
  &= 
  g^2 
  \left( {e^{-i\pi \Delta_\phi/2} \over 2^{\Delta_\phi+1} \pi R^{1/2}\Gamma(\Delta_\phi)}\right)^4
 \int_{AdS_3} dY_1 \int_{AdS_3} dY_2\,\notag\\
 &\times
\left( \prod_{a=1}^4  \int_0^\infty d\omega_a \omega_a^{\Delta_\phi-1}\right)
 e^{-i Y_1\cdot (\omega_1 P_1 +\omega_2 P_2)-i Y_2 \cdot(\omega_3 P_3 +\omega_4 P_4) -\epsilon \sum_a \omega_a }
  \Pi_\Delta(Y_1,Y_2)
 \,.
\end{align}
Let us now approach the bulk point singularity by tuning the boundary points to be close to the configuration \eqref{bulkpt1} and \eqref{bulkpt2}.  We will choose $P_{1,2}$ to be $(0,-q_{1,2}^\mu)$ and while $P_{3,4}$ to be $(\delta , q_{3,4}^\mu)$.  If we also choose $q_a^\mu$ to be such that \eqref{bulkpt2} is obeyed, then we reach the bulk point singularity configuration as $\delta\to0$. 
In this limit we expect  the contribution of the integral in \eqref{I4} to be dominated by $Y_1, Y_2$ close to  $Y_0=(R,0,0,0)$, which can then be approximated by integrals over $\mathbb{R}^{1,d}$.    To be more explicit, near $Y_0$, we can parametrize a bulk point $Y$ as $Y^I \simeq (R+{X^\mu X_\mu\over 2R} , X^\mu)$, where
every component of $X^\mu$ is much less that $R$.  Equivalently, we can take $R\to\infty$ and allow  $X^\mu \in \mathbb{R}^{1,2}$ to be integrated to infinity in \eqref{I4}. 

In this limit, the Witten diagram becomes
\begin{align}\label{finalI4}
I_4 &\simeq  g^2\left( {e^{-i\pi \Delta_\phi/2} \over 2^{\Delta_\phi+1} \pi R^{1/2}\Gamma(\Delta_\phi)}\right)^4
\left( \prod_{a=1}^4  \int_0^\infty d\omega_a \omega_a^{\Delta_\phi-1}\right)
\notag\\
&\times \int_{\mathbb{R}^{1,2}}d^3X_1\int_{\mathbb{R}^{1,2}}d^3X_2
e^{i X_1 \cdot (\omega_1 q_1 +\omega_2 q_2) - i X_2\cdot (\omega_3 q_3 +\omega_4q_4)  +  i R  (\omega_3+\omega_4)\delta -\epsilon\sum_a\omega_a}
\Pi_\Delta(Y_1,Y_2)\,.
\end{align}
If we keep the intermediate conformal dimension $\Delta$ finite as taking $R\to \infty$, then the bulk-to-bulk propagator $\Pi_\Delta(Y_1,Y_2)$ is approximated by the \textit{massless} Feynman scalar propagator in $\mathbb{R}^{1,2}$, i.e. $\Pi_\Delta\simeq {1\over 4\pi} {1\over \sqrt{(X_1-X_2)^2 +i \epsilon}}$.  The $\omega_a$ integrals then give rise to a singularity $1/\delta^{4\Delta_\phi -5}$ as $\delta\to0$. This is indeed the singularity computed in \cite{Gary:2009ae} using the explicit expression for $I_4$ in terms of the $D$-function for special values of $\Delta$.  In the strict $\delta=0$ case, the above integral is identical to the massless scalar exchange amplitude in $\mathbb{R}^{1,2}$ written in the massless conformal basis.  As discussed in Section \ref{sec:scalarexchange}, the change of basis integral \eqref{main4pt} to the conformal basis suffers from an UV divergence (for sufficiently positive $\Delta_\phi$).  Now we have an alternative understanding of this singularity: it comes from  the bulk point singularity of the same interaction in $AdS_3$. 

If instead we scale the intermediate conformal dimension $\Delta$ to infinite at the same rate as sending $R\to\infty$, while holding the ratio $m\equiv \Delta/R $ fixed,\footnote{This flat space limit was considered, for example, in \cite{Penedones:2010ue,Paulos:2016fap}.} then the $AdS_3$ bulk-to-bulk propagator is approximated by the \textit{massive} scalar Feynman propagator in $\mathbb{R}^{1,2}$:
\begin{align}
\Pi_\Delta(Y_1,Y_2) \simeq {1\over 4\pi }{1\over \sqrt{(X_1-X_2)^2+i\epsilon}}e^{- m \sqrt{(X_1-X_2)^2+i\epsilon}}\,.
\end{align}
The second line of \eqref{finalI4} is nothing but the Fourier transform from position space to momentum space for the flat space amplitude ${1\over s-m^2 + i\epsilon} \delta^{(3)}(\sum_a k_a)$ with null momenta $k_a = \pm \omega_aq_a$. 
Thus the scalar exchange Witten diagram \eqref{finalI4} in this double scaling limit  exactly reproduces the same Feynman diagram in $\mathbb{R}^{1,2}$ written in the massless conformal basis, which we computed in Section \ref{sec:scalarexchange}:
\begin{align}
I_4\propto \left( \prod_{a=1}^4  \int_0^\infty d\omega_a \omega_a^{\Delta_\phi-1}\right)
{\delta^{(3)}(\omega_1q_1^\mu +\omega_2 q_2^\mu -\omega_3q_3^\mu -\omega_4q_4^\mu)\over s-m^2 + i\epsilon} \,,~~~~s=-(\omega_1q_1+\omega_2q_2)^2\,.
\end{align} 
The infinite intermediate conformal dimension $\Delta$ limit regulates the $\omega_a$ integrals near the would-be bulk point singularity, while it damps the contribution from the $AdS$ integrals away from the point $Y_0$.

From this perspective, the (2+1)-dimensional flat space amplitudes written in the massless conformal basis $f(z)$ should perhaps be interpreted as \textit{two-dimensional Lorentzian} conformal correlators, restricted to the bulk point singularity configuration $z=\bar z $.  In Appendix \ref{app:2d}, we present a $2d$ Euclidean correlator whose Lorentzian versions, when restricted to $z=\bar z$, are exactly the (2+1)-dimensional tree-level exchange amplitudes in the conformal basis $f_{12\leftrightarrow 34}(z),f_{13\leftrightarrow 24}(z),f_{14\leftrightarrow 23}(z)$.  We leave the physical origin of this $2d$ Euclidean correlator for future investigation.

 \section*{Acknowledgements} 
We are grateful to N. Arkani-Hamed,  C.-M. Chang,  Y.-t. Huang, S. Komatsu,  Y.-H. Lin, J. Maldacena, P. Mitra, H. Ooguri,   M. Spradlin, D. Stanford,  A. Volovich, Y. Wang, E. Witten,   and especially to 
Sabrina Pasterski and Andy Strominger for interesting discussions. 
 We thank S. Pasterski, D. Simmons-Duffin, A. Strominger, and A. Zhiboedov for comments on a draft. 
HTL is grateful to ICTP-SAIFR for their hospitality.
SHS is grateful to National Taiwan University and the Aspen Center for Physics for their hospitality.  
HTL is supported by a Croucher Scholarship for Doctoral Study and a Centennial Fellowship from Princeton University.   
SHS is supported by the Zurich Insurance Company Membership and the National Science Foundation grant
PHY-1314311.

\appendix

\section{Two-Dimensional Crossing Symmetric Four-Point Functions  with Positive Block Decompositions}\label{app:2d}

In this appendix we present a $2d$ Euclidean correlator $f^{2d}(z,\bar z)$ whose Lorentzian versions, obtained via different analytic continuations in $z,\bar z$, are the (2+1)-dimensional amplitudes from three crossing channels $f_{12\leftrightarrow 34}(z),f_{13\leftrightarrow 24}(z),f_{14\leftrightarrow 23}(z)$.  We checked to high orders that this $2d$ correlator admits a positive $SL(2,\mathbb{C})$ block decomposition. For a special value of the external conformal dimension $\Delta_\phi$, $f^{2d}(z,\bar z)$  reduces to the four-point function in the $2d$ free boson theory. 

\subsection{A Two-Dimensional Extension}

We now present an observation that in the example of tree-level  exchange diagram,  the correlators in the three crossing channels $ f_{12\leftrightarrow 34}, f_{13\leftrightarrow 24},f_{14\leftrightarrow 23}$ are restrictions of different analytic continuations of a single 2$d$ Euclidean correlator $f^{2d}(z,\bar z)$.  

Let the two cross ratios $z,\bar z$ of a $2d$ four-point function bedefined as
\begin{align}
z  = {z_{12}z_{34}\over z_{13}z_{24}}\,,~~~~ \bar z  = {\bar z_{12}\bar z_{34}\over\bar z_{13}\bar z_{24}}\,.
\end{align}
In the Euclidean signature, $z,\bar z$ are complex conjugated to each other, i.e. $\bar z= z^*$.  On the other hand, in the Lorentzian signature, $z,\bar z$ are two independent real variables.  
Starting from an Euclidean $2d$ four-point function,  its Lorentzian version can be obtained by analytic continuing the cross ratios $z,\bar z$ independently.  The precise analytic continuation depends on the  time ordering between the four operators in the Lorentzian spacetime.


 If we place  operators 1 and 2 in the past while 3 and 4 in the future, this corresponds to the following analytic continuation in $z,\bar z$ (see, for example, \cite{Hartman:2015lfa,Maldacena:2015iua} for explanations of this analytic continuation):
\begin{align}\label{ACs}
12\to 34:~(z-1)\to e^{2\pi i }(z-1)\,,~~~~{1\over \bar z} \to e^{2\pi i }{1\over \bar z}\,.
\end{align}
If instead we want to have 1 and 3 in the past while 2 and 4 in the future,  we should analytically continue the cross ratios as
\begin{align}\label{ACt}
13 \to 24:~ z\to e^{2\pi i} z\,,~~~~(\bar z-1)\to e^{2\pi i}(\bar z-1)\,.
\end{align}
In addition, due to the overall factor ${1\over |z_{12}|^{2\Delta_\phi}|z_{34}|^{2\Delta_\phi}}$, one needs to further multiply the four-point function $f(z, \bar z)$ by $e^{-2\pi i \Delta_\phi}$.  This is because $|z_{12}|^2$ crosses the lightcone in the analytic continuation.
Finally, if we place 1 and 4 in the past, 2 and 3 in the future, we analytically continue the cross ratios as
\begin{align}\label{ACu}
14 \to 23:~ \frac 1z\to e^{2\pi i }\frac 1z\,,~~~~\bar z \to e^{2\pi i } \bar z\,.
\end{align}
In addition, we need to multiply the four-point function by a phase $e^{-2\pi i \Delta_\phi}$.

Now consider the following two-dimensional Euclidean four-point function
\begin{align}\label{2d4pt}
& {\mathcal{A}}(z,\bar z) = {1\over |z_{12}|^{2\Delta_\phi} |z_{34}|^{2\Delta_\phi}}  f^{2d}(z,\bar z)\,,\\
&  f^{2d}(z,\bar z) =
\mathcal{N}_{\Delta_\phi} e^{i\pi (2\Delta_\phi -\frac 12)}
 {(z\bar z)^{\frac12}\over [(1-z)(1-\bar z)]^{\frac 14}}\left[
1+(z\bar z)^{\Delta_\phi-\frac 34 }  + \left({z\bar z\over (1-z)(1-\bar z)}\right)^{\Delta_\phi-\frac 34}
\right] \,.
\end{align}
This $2d$ Euclidean correlator satisfies the crossing equation \eqref{2dcrossing} and admits a positive $SL(2,\mathbb{C})$ block decomposition as discussed in the next subsection.  
If we first analytically continue $f^{2d}(z,\bar z)$ to the Lorentzian regime as \eqref{ACs}, \eqref{ACt}, \eqref{ACu}, and then restrict to $z=\bar z$, we exactly reproduce the (2+1)-dimensional tree-level exchange amplitudes $ f_{12\leftrightarrow 34}, f_{13\leftrightarrow 24}, f_{14\leftrightarrow 23}$ \eqref{phi3s}, \eqref{phi3t}, \eqref{phi3u} in three different crossing channels, respectively:
\begin{align}
\begin{split}
& f_{12\leftrightarrow 34}(z) =
 f^{2d}(z,\bar z)\Big|_{(z-1)\to e^{2\pi i }(z-1), \frac {1}{\bar z}\to e^{2\pi i } {1\over \bar z}}
\, \Big|_{z=\bar z}\,,\\
& f_{13\leftrightarrow 24}(z) =e^{-2\pi i \Delta_\phi} f^{2d}(z,\bar z)\Big|_{z\to e^{2\pi i }z, (\bar z-1)\to e^{2\pi i } (\bar z-1)}
\, \Big|_{z=\bar z} \,,\\
& f_{14\leftrightarrow 23}(z) =e^{-2\pi i \Delta_\phi}  f^{2d}(z,\bar z)\Big|_{\frac 1z\to e^{2\pi i }\frac 1z, \bar z\to e^{2\pi i }\bar z}  \,  \Big|_{z=\bar z}\,,
\end{split}
\end{align}
where the phase $e^{-2\pi i \Delta_\phi}$ comes from the prefactor ${1\over |z_{12}|^{2\Delta_{\phi}} |z_{34}|^{2\Delta_\phi}}$ in our definition of $\tilde f(z,\bar z)$ in relation to the full four-point function. This implies that the one-dimensional four-point functions are restrictions of $f(z,\bar z)$ to the Lorentzian configurations  similar to Figure \ref{fig:bulkpoint}.


\subsection{Positive Conformal Block Decompositions}

In fact, the Euclidean $2d$ four-point function $f^{2d}(z,\bar z)$ \eqref{2d4pt} belongs to  a larger family of solutions to the crossing equation.  In this section we provide numerical evidence that this family of crossing solutions have positive conformal block decompositions. 

Let us consider a generalization of the $2d$ scalar four-point function \eqref{2d4pt} considered above:
\begin{align}
f_{\Delta_\phi,b}(z,\bar z) =  {( z\bar z)^{\Delta_\phi-\frac b2}  \over ((1-z)(1-\bar z))^{b\over2} }\left[ \,
(z\bar z)^{-\Delta_\phi+{3b\over2}}  +1+ ((1-z)(1-\bar z))^{-\Delta_\phi +{3b\over2}}\,
\right]\,.
\end{align}
This four-point function comes with a two-parameter family, labeled by the external operator dimension $\Delta_\phi$ (which will be taken to be positive in this appendix)  and a real parameter $b$.  
The four-point function $f_{\Delta_\phi,b}$ is crossing symmetric for all $\Delta_\phi,b$:
\begin{align}\label{2dcrossing}
f_{\Delta_\phi,b}(z,\bar z) = \left({ z\bar z \over (1-z)(1-\bar z) }\right)^{\Delta_\phi} f_{\Delta_\phi,b}(1-z,1-\bar z) = (z\bar z)^{\Delta_\phi} f_{\Delta_\phi,b}(1/z,1/\bar z)\,.
\end{align}

For  general $\Delta_\phi$ and $b$, the operator spectrum in the OPE channel  $z,\bar z\to 0$ can be separated into towers of ``single-trace" and the ``double-trace" operators:
\begin{align}\label{spectrum}
\begin{split}
&\text{Single-trace}:\,~~ \Delta = 2b +  \ell +4m\,,~~~~~~\,~~~~m\in \mathbb{Z}_{\ge0}\,,\\
&\text{Double-trace}:~ \Delta = 2\Delta_\phi -b +\ell +2m \,,~~~m\in \mathbb{Z}_{\ge0}\,,
\end{split}
\end{align}
where $\ell \in 2\mathbb{Z}_{\ge0}$ is the spin.  We distinguish a double-trace operator from a single-trace one by the dependence of their conformal dimensions on $\Delta_\phi$.   
The conformal block expansion of the four-point function can be separated into single- and double-trace operators:
\begin{align}\label{2dCBD}
f_{\Delta_\phi,b}(z, \bar z) =  \sum_{ \substack{ \ell=0,\\\text{even}}}^\infty\sum_{m=0}^\infty
 c_{\ell+4m,\ell} \, G_{2b+\ell +4m}^{(\ell)}(z,\bar z)
+\sum_{ \substack{ \ell=0,\\\text{even}}}^\infty\sum_{m=0}^\infty
 c^d_{\ell+2m,\ell} \, G_{2\Delta_\phi-b+\ell+2m}^{(\ell)}(z,\bar z)\,,
\end{align}
where the $SL(2,\mathbb{C})$ block with identical external scalar primaries of dimension $\Delta_\phi$ and intermediate conformal dimension $\Delta$ and spin $\ell\ge0$ is \cite{Ferrara:1974ny,Dolan:2000ut,Dolan:2003hv,Dolan:2011dv}
\begin{align}
G^{(\ell)}_\Delta(z,\bar z) =&{1\over 1+\delta_{\ell,0}}\Big[
 z^{\frac12 (\Delta+\ell)} \, \bar z^{\frac12 (\Delta-\ell)}\,   \,_2F_1\left({ \Delta+\ell\over 2},{\Delta+\ell\over2} ;\Delta+\ell;z\right)\notag\\
&
\times\,_2F_1\left({\Delta-\ell\over2},{\Delta-\ell\over2};\Delta-\ell;\bar z\right)
+(z\leftrightarrow \bar z)\,\Big]
\,.
\end{align} 
The first few coefficients are
\begin{align}
&c_{0,0}=1\,,~~c_{4,0} = {b^4 \over 64 (1+2b)^2} \,,~~ 
c_{2,2}={b^2\over 8(2b+1)}\,,~~
c_{6,2}={b^4 (b+2)^2 \over1024 (2b+1)(2b+3)(2b+5)}\,,~~\notag\\
&c_{8,0} = {b^4 (b+2)^4\over 16384(2b+3)^2 (2b+5)^2}\,,
~~c_{4,4}= {b^2(b+2)^2 \over 128 (2b+3) (2b+5)}\,,\cdots\,,\\
&c^d_{0,0}=2\,,~~c^d_{2,0}=\frac 18(3b-2\Delta_\phi)^2\,,~~
c^d_{4,0}= \frac{\left(9 b^3-10 b^2 (3 \Delta_\phi +1)+4 b \Delta_\phi  (7 \Delta_\phi +4)-8 \Delta_ \phi ^2 (\Delta_ \phi +1)\right)^2}{512 (b-2 \Delta _\phi -1)^2}\,,\notag\\
& c^d_{2,2}=\frac{9 b^3-10 b^2 (3 \Delta_ \phi +1)+4 b \Delta _\phi  (7 \Delta _\phi +4)-8 \Delta _\phi ^2 (\Delta_ \phi +1)}{16 (b-2 \Delta_ \phi -1)}\,,\cdots.
\end{align}
We numerically checked that these conformal block coefficients are non-negative if
\begin{align}\label{bbound}
2\Delta_\phi \ge b\ge 0\,.
\end{align}
Note that this constraint also arises from requiring  the leading single-trace and double-trace operators \eqref{spectrum} to have positive conformal dimensions.

\begin{figure}[h!]
\begin{center}
\includegraphics[width=.6\textwidth]{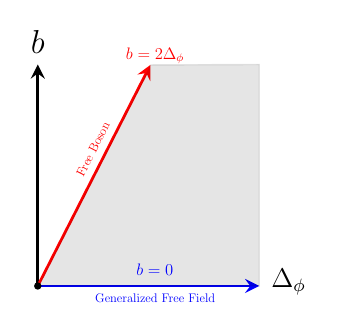}
\end{center}
\caption{The two-parameter space for the solution $f_{\Delta_\phi,b}(z,\bar z)$ to the $2d$ crossing equation.  In the shaded region between $0\le b\le 2\Delta_\phi$ we find numerical evidence that the four-point function $f_{\Delta_\phi,b}(z,\bar z)$ has a non-negative expansion on the $SL(2,\mathbb{C})$ blocks.  At the two boundaries $b=0$ and $b=2\Delta_\phi$, $f_{\Delta_\phi,b}(z,\bar z)$ reduces to the four-point functions in the $2d$ generalized free field theory and that in the free boson theory, respectively.}\label{fig:2d4pt}
\end{figure}

The intermediate spectrum contains the identity operator only if $b=0$ or $b=2\Delta_\phi$, which are the lower and upper bounds of $b$ \eqref{bbound}.   
In fact, at these values of $b$ the  four-point function $f_{\Delta_\phi,b}(z,\bar z)$ reduces to known examples:
\begin{itemize}
\item $b=0$. In this case  $f_{\Delta_\phi,b=0}(z,\bar z)$ reduces to the four-point function in the $2d$ generalized free field theory:
\begin{align}
f_{\Delta_\phi,b=0}(z,\bar z) =  
1+ 
(z\bar z)^{\Delta_\phi}  
+ \left( {z\bar z \over (1-z)(1-\bar z)} \right)^{\Delta_\phi}\,.
\end{align}
\item $b=2\Delta_\phi$. In this case $f_{\Delta_\phi , b=2\Delta_\phi}$ reduces to the four-point function of scalar primaries $\cos(\sqrt{\Delta_\phi} X)$ in the $2d$ free boson theory:\footnote{We thank Ying-Hsuan Lin for discussions about this point.}
\begin{align}
f_{\Delta_\phi,b=2\Delta_\phi}(z,\bar z)
= {1\over 2((1-z)(1-\bar z))^{\Delta_\phi} }
\Big[ \,(z\bar z) ^{2\Delta_\phi}  + 1 + ((1-z)(1-\bar z))^{2\Delta_\phi}
\,\Big]\,.
\end{align}
Here we have normalized the four-point function by a factor of $1/2$ so that the identity channel comes with unit OPE coefficient.  The radius of the free boson does not affect the four-point function as long as the operator $\cos(\sqrt{\Delta_\phi}X)$ exists in the spectrum of primaries. 
\end{itemize}

To conclude,  $f_{\Delta_\phi,b}$ provides an interpolation between the four-point function in the $2d$ generalized free field theory and that in the free boson theory (see  Figure \ref{fig:2d4pt}).  Even though $f_{\Delta_\phi,b}$ has no identity channel away from the two limiting cases, we find numerical evidence that it admits a non-negative $SL(2,\mathbb{C})$ block decomposition.

\bibliography{3dscalar}{}
\bibliographystyle{utphys}
\end{document}